\documentstyle[epsf,twoside,fleqn,espcrc2]{article}

\def\g#1{{\scriptstyle (\! #1 \! )}}
\def\Sg#1{{\scriptstyle [\! #1 \! ]}}

\def\gg#1{{\scriptscriptstyle (\! #1 \! )}}

\def\IUg[#1]{ {\mathsf{I}}^{\scriptstyle [\! #1 \! ]}{}}
\def\IDg[#1]{ {\mathsf{I}}_{\scriptstyle [\! #1 \! ]}{}}
\def\IUgg[#1]{{\mathsf{I}}^{\scriptscriptstyle [\! #1 \! ]}{}}
\def\IDgg[#1]{{\mathsf{I}}_{\scriptscriptstyle [\! #1 \! ]}{}}
\def\xp[#1]{{{\mathbf{x}}\!+\!#1}}
\def\xm[#1]{{{\mathbf{x}}\!-\!#1}}
\def\yp[#1]{{{\mathbf{y}}\!+\!#1}}
\def\ym[#1]{{{\mathbf{y}}\!-\!#1}}

\def\Rgen[#1]#2#3{{ 
        D^{ {\scriptstyle (\! #1 \! )}#2}_{~~~#3} }}
\def\gUP[#1]#2{{   g_{{[\! #1 \! ]}}^{{#2 }} }}
\def\gDOWN[#1]#2{{ g^{{[\! #1 \! ]}}_{{#2 }} }}
\def\ThreeJUP[#1]#2{  \left|{}_{[\!#1\!]}^{#2} \right| }
\def\ThreeJDOWN[#1]#2{\left|{}^{[\!#1\!]}_{#2} \right|}

\newcommand{\beq}{\begin{eqnarray}}
\newcommand{\eeq}{\end{eqnarray}}

\newcommand{\np}{Nucl.Phys.\ }
\newcommand{\pl}{Phys.Lett.\ }
\newcommand{\pr}{Phys.Rev.\ }

\newcommand{\be}{\begin{equation}}
\newcommand{\ee}{\end{equation}}

\def\np#1#2#3{Nucl.\ Phys.\ B#1 (19#3) #2}
\def\pl#1#2#3{Phys.\ Lett.\ #1B (19#3) #2}
\def\pr#1#2#3{Phys.\ Rev.\ D #1 (19#3) #2}

\def\rmp#1#2#3{Rev.\ Mod.\ Phys.\ #1 (19#3) #2}

\newcommand{\AmS}{{\protect\the\textfont2
  A\kern-.1667em\lower.5ex\hbox{M}\kern-.125emS}}

\title{Hamiltonian $LGT$ in the complete Fourier analysis basis.}

\author{\
G. Burgio\address{Dipartimento di Fisica, Universit\`a di Parma
        and INFN, Gruppo Collegato di Parma, Parma, Italy}, 
R. De Pietri\address{Centre de Physique Th\'eorique CNRS, 
        Case 907 Campus de Luminy, F-13288 Marseille Cedex 9, France},
H. A. Morales-T\'ecotl\address{Departamento de F\'\i sica, 
        Universidad Aut\'onoma Metropolitana Iztapalapa, 
        A. Postal 55-534, 09340   M\'exico, D.F.},
L. F. Urrutia\address{Departamento de F\'\i sica de Altas Energ\'\i as, 
        Instituto de Ciencias Nucleares, Universidad Nacional Aut\'onoma 
        de M\'exico, A. Postal 70-543, 04510 M\'exico D.F.}
and J. D. Vergara$^{\mathrm d}$.
}

\begin{document}

\begin{abstract}
The main problem in the Hamiltonian formulation of Lattice Gauge
Theories is the determination of an appropriate basis avoiding the
over-completeness arising from Mandelstam relations.  We short-cut
this problem using Harmonic analysis on Lie-Groups and intertwining
operators formalism to explicitly construct a basis of the Hilbert
space.  Our analysis is based only on properties of the tensor category
of Lie-Group representations.  The Hamiltonian of such theories is
calculated yielding a sparse matrix whose spectrum and eigenstates
could be exactly derived as functions of the coupling $g^2$.
\vskip 2mm
Preprint UPRF-99-12, September 1999. 
To appear in Nucl.\ Phys.\ Proc.\ Suppl., LATTICE99.
\end{abstract}

\maketitle

\section{HAMILTONIAN LATTICE GAUGE THEORY}
\protect\label{sec:HLGT}

The Hamiltonian formalism for lattice gauge theories (LGT) 
in $(d\!+\!1)$ dimensions \cite{Kogut:1975} is constructed 
associating gauge field variables $U_k({\mathbf{x}})\in G$  
to each link $({\mathbf{x}},{\mathbf{x}}+a{\mathbf{e}}_k)$ of 
a hypercubic periodic lattice of period $a L$.  
The corresponding Hilbert space ${\mathcal{H}}$ is defined by the 
gauge invariant square integrable functions
$\psi(U)=\psi(U^\gamma)=\psi(\{ U_k^\gamma(\mathbf{x}) \})$ 
on the tensor product of $d \cdot L^d$ copies of the 
gauge group $G$. Gauge transformations act as $U_k({\mathbf{x}}) 
\longrightarrow U_k^\gamma({\mathbf{x}}) 
= \gamma^{-1}({\mathbf{x}}+a{\mathbf{e}}_k)
  U_k({\mathbf{x}}) \gamma({\mathbf{x}})$.
The variables conjugated to $U_k(\mathbf{x})$ are the
outgoing/ingoing electric fields $E^\alpha_{\pm\! k}({\mathbf{x}})$
from the lattice point ${\mathbf{x}}$ in the directions 
$ {\mathbf{e}}_{\pm\!k}$. 

The standard Hamiltonian operator is 
\begin{equation} \label{def:HAM}
\hat{H} =       
        \frac{g^2}{2a^{d\!-\!2}} \sum_{{\mathbf{x}},k} 
        q_{\alpha\beta} E^\alpha_k({\mathbf{x}})E^\beta_k({\mathbf{x}})
     + \sum_P V(U_P), 
\end{equation}
where $q_{\alpha\beta}$ is the Cartan metric, the sum over $P$ 
ranges over all unoriented plaquettes, $U_P$ is the 
plaquette variable and
\begin{equation}
  V(U_P) = \frac{a^{d\!-\!4}}{g^2} \left[1 - 
           \frac{U_P + U_P^*}{2\mathrm{dim}(U)} \right] 
  ~~.
\end{equation}
Such choice is not unique since the only condition on the magnetic 
term potential is $V(U_P)\simeq\frac{a^4}{2} {\rm Tr}[F_P^2]$.

\section{THE SPIN NETWORK BASIS}

A classical result of representation theory
gives a nice way of constructing a basis of LGT Hilbert space.
In fact, the set  ${\mathcal{RG}} = \{  {\mathcal{R}}^{j} ~| j\in J[G]  \}$ 
of all the unitary inequivalent representations of 
a compact group $G$ is numerable and all the representations
${\mathcal{R}}^{j}$ are finite dimensional. Choosing an orthonormal basis for 
each representation ${\mathcal{R}}^{j}$, the matrix elements 
$\Rgen[j]{\alpha}{\beta}(U)$ 
($\alpha,\beta=1,\ldots,\mathrm{dim}({\mathcal{H}}^j)$) 
of all the representations ${\mathcal{R}}^{j}$ 
are a numerable orthonormal basis of ${\mathcal{L}}^2[G,dU]$.
This result, known as the Peter-Weyl theorem \cite{BookGroup}, implies 
that each vector of $\mathcal{H}$ can be written as
\def\SHORTCUTforC{
{c^{\g{j_\gg{1}\cdots j_\gg{N_{lk}}}
    \beta_\gg{1}\cdots\beta_\gg{N_{lk}}
}_{~~~~~~~~~~~\alpha_\gg{1}\cdots\alpha_\gg{N_{lk}}} }}
\def\SHORTCUTforCbar{
{c^{\g{j_\gg{1}\cdots j_\gg{N_{lk}}}
    \bar{\beta}_\gg{1}\cdots\bar{\beta}_\gg{N_{lk}}
}_{~~~~~~~~~~~\bar{\alpha}_\gg{1}\cdots\bar{\alpha}_\gg{N_{lk}}} }}
\begin{eqnarray} \label{eq:vHaux}
&&  \psi(U) = \prod_{\mathbf{x}} \prod_{k=1}^{d} 
     \sum_{j_{\mathbf{x}}^k \in J[G]} 
     \sum_{\alpha_{\mathbf{x}}^k,\beta_{\mathbf{x}}^k=1}^{
                         {{\mathrm{dim}}(j_{\mathbf{x}}^k)}}
  \bigg[ \\
&& \qquad
     ~\Rgen[j_{\mathbf{x}}^k]{\alpha_{\mathbf{x}}^k}{\beta_{\mathbf{x}}^k}(U) 
    \times 
    \SHORTCUTforC
  \bigg],
\nonumber
\end{eqnarray}
where only gauge invariant combinations should be taken into account.

The implementation of gauge invariance turns into a set of
constraints on the coefficients $c$. In particular the $c$'s should
factorize in products of group invariant tensors associated to the
different lattice sites ${\mathbf{x}}$.

The concept of invariant tensor is better expressed by the notion of
intertwining operators.  By definition, an operator 
${\mathsf I}$ connecting the Hilbert space of two 
representations, $\mathcal{R}$ and
${\mathcal{R}}'$, is an intertwining operator if ${\mathsf{I}} \cdot
T(U) = T'(U) \cdot {\mathsf{I}}$, for every $U$ in $G$.  
The set of all intertwining operators 
${\mathcal{I}}({\mathcal{R}},{\mathcal{R}}')$ is a vector subspace 
of all the linear operators connecting the Hilbert space of the two 
representations ${\mathcal{R}}$ and ${\mathcal{R}}'$.
This concept gives 
the coordinate free definition of the generalized Clebsh-Gordan
coefficients of Yutsis-Levinson-Vanagas
\cite{Yutsis:1962} which are the matrix elements of these operators on
the chosen basis. The integral of the product of $K$ 
representations decomposes according to 
\begin{equation}
 \int\!\! dU 
  \prod_{k=1}^K  \Rgen[j_k]{\alpha_k}{\beta_k}(U)
\!=\! \sum_{\pi}
  \frac{\IUg[\pi]{}^{(j_1\ldots j_K)}_{\beta_1\ldots\beta_K}
        \IDg[\pi]{}_{(j_1\ldots j_K)}^{\alpha_1\ldots\alpha_K}
      }{\IUg[\pi]{}^{(j_1\ldots j_K)}_{\gamma_1\ldots\gamma_K}
        \IDg[\pi]{}_{(j_1\ldots j_K)}^{\gamma_1\ldots\gamma_K}
       },
\end{equation}
where $\IDg[\pi]{}_{(j_1\ldots j_K)}\in {\mathcal{I}}({\mathcal{R}}^{j_1}
\otimes\ldots\otimes{\mathcal{R}}^{j_K},\emptyset)$, 
$\IUg[\pi]{}^{(j_1\ldots j_K)}$ is 
its adjoint and the sum is extended over a complete 
orthogonal basis of ${\mathcal{I}}({\mathcal{R}}^{j_1}
\otimes\ldots\otimes{\mathcal{R}}^{j_K},\emptyset)$.

Summarizing, Peter-Weyl theorem and gauge invariance 
leads to the  {\it spin network} basis elements
\begin{eqnarray} 
&& \psi_{\vec{\jmath},\mathbf{\vec{\pi}}}(U) 
   = \prod_{{\mathbf{x}}}~ 
     \prod_{k=1}^{d}             
              \sum_{\alpha_{{\mathbf{x}}}^k,\beta_{{\mathbf{x}}}^k=1}^{
                   {{\mathrm{dim}}(j_{{\mathbf{x}}}^k)}} 
     \bigg[
\label{eq:spinNET} \\
&&\qquad    
   ~\Rgen[j_{\mathbf{x}}^k ]{{\alpha}_{\mathbf{x}}^k}{
         {\beta}_{\mathbf{x}}^k}(U_k({\mathbf{x}}))  
    \cdot
    \mathsf{I}_{{\mathbf{x}}}^{\Sg{{\mathbf{\pi}}_{\mathbf{x}}}}
    {}^\g{j_{\xm[d]}^1,\ldots,j_{\xm[d]}^d}_{
        \alpha_{\xm[1]}^1,\ldots,\alpha_{\xm[d]}^d}
    {}_\g{j_{\mathbf{x}}^1,\ldots,j_{\mathbf{x}}^d}^{
        \beta_{\mathbf{x}}^1,\ldots,\beta_{\mathbf{x}}^d}
    \bigg]~.
\nonumber 
\end{eqnarray}


\section{MATRIX ELEMENTS OF THE HAMILTONIAN OPERATOR}

Computing the action of the Hamiltonian operator 
(\ref{def:HAM}) on the {\it spin-networks basis} simply reduces 
to tracing intertwining operators.  In fact, the basis vectors 
(\ref{eq:spinNET}) are eigenstates of the  kinetic term, while the 
potential (magnetic) term is realized as a multiplicative operator.
Explicitly
\begin{eqnarray} \label{eq:HspinNET}
   \langle \vec{\jmath}~',{\mathbf{\vec{\pi}}'}| 
   \hat{H} | \vec{\jmath},{\mathbf{\vec{\pi}}}\rangle =  
  \frac{- a^{d\!-\!4}}{2\ g^2\ {\rm dim}(U)}
  \sum_{\mathbf{y}} \sum_{r<s=1..d} \times
\nonumber
\\ \nonumber ~
   \times \bigg(
   \langle \vec{\jmath}~',{\mathbf{\vec{\pi}}'}| 
   U_{{\mathbf{y}},r,s} | \vec{\jmath},{\mathbf{\vec{\pi}}}\rangle
  +\langle \vec{\jmath},{\mathbf{\vec{\pi}}}| 
   U_{{\mathbf{y}},r,s} | \vec{\jmath}~',{\mathbf{\vec{\pi}}'}\rangle
\bigg)
\\ \nonumber ~ 
+ \left( \frac{g^2}{2\ a^{d\!-\!2}} 
     \sum_{\mathbf{x}}\sum_{k=1}^{d}
              C_2[j_{\mathbf{x}}^2]  ~
     +\frac{a^{d\!-\!4}}{g^2} N_P \right) 
        \delta_{\vec{\jmath}}^{\vec{\jmath}'}
        \delta_{\mathbf{\vec{\pi}}}^{{\mathbf{\vec{\pi}}}'},
\nonumber
\end{eqnarray}
where the only non diagonal terms are given by the expectation 
values of the plaquette operator. These are given in equation (26) of  
Ref.\cite{BurgioEtAl:1999} as traces of intertwining 
operators.  In this way the computation amounts to the evaluation
of specific Wigner's $nJ$-symbols, that further reduce to 
$6J$-symbols only.

An important property of the Hamiltonian matrix in 
the {\it spin-network} basis is that it is sparse.

\begin{figure}[htb]\protect\label{fig}
\def\BasicBlockTwoDim#1#2#3#4#5{{
\mbox{{\setlength{\unitlength}{1pt}
\begin{picture}(70,70)
   \put(35,35){\circle{25}}
   \put(30,30){\line( 1,1){10}}
   \thicklines 
   \put( 0,35){\line(6,-1){30}}   \put(70,35){\line(-6, 1){30}}   
   \put(35, 0){\line(-1,6){5}}    \put(35,70){\line( 1,-6){5}}   
   \thinlines 
   \put(30,30){\circle*{3}}      \put(40,40){\circle*{3}}
   \thinlines 
   \put( 7,38){$\scriptstyle #1$}
   \put(36,10){$\scriptstyle #2$}
   \put(33,29){$\scriptstyle #3$}
   \put(52,42){$\scriptstyle #4$}
   \put(40,53){$\scriptstyle #5$} 
\end{picture}}}  }}
\mbox{\setlength{\unitlength}{1pt}
\begin{picture}(140,140)
   \put( 0, 0){\BasicBlockTwoDim{}{}{
               \pi_{\mathbf{x}}^1}{j_{\mathbf{x}}^1}{j_{\mathbf{x}}^2}}
   \put(70, 0){\BasicBlockTwoDim{}{j^2_{\xp[1\!-\!2]}}{
                 \pi^1_{\xp[1]}}{j^1_{\xp[1]}}{j^2_{\xp[1]}}}
   \put( 0,70){\BasicBlockTwoDim{j_{\xm[1\!+\!2]}^1}{}{
                      \pi^1_{\xp[2]}}{j^1_{\xp[2]}}{j^2_{\xp[2]}}}
   \put(70,70){\BasicBlockTwoDim{}{}{\pi^1_{\xp[1\!+\!2]}}{}{}}
\end{picture}}
\caption{A spin network in the case of a 2 dimensional lattice is 
  parametrized by an irreducible representation associated to each 
  link {\protect{$j_{\mathbf{x}}^{1,2}$}} and an 
  irreducible representation \protect{$\pi^1_{\mathbf{x}}$}
  parametrizing the irreducible tensor associated to each lattice 
  site.}
\label{fig:VERTEX}
\end{figure}
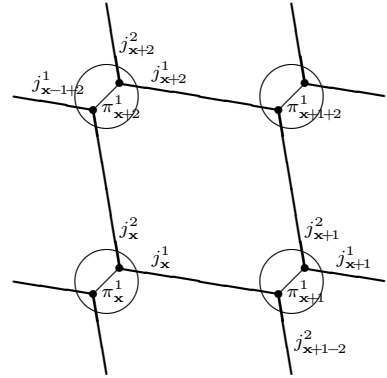
For example, in  a two dimensional lattice and gauge group $SU(2)$ 
the spin-network basis elements are characterized  
by three half integer associated to each lattice site 
(see Fig. 1).  
The matrix elements of the plaquette $U_{{\mathbf{y}},1,2}$
are different from zero only if all the six primed and un-primed  
$j_{\mathbf{x}}^1$, $j_{\mathbf{x}}^2$, 
$\pi^1_{\xp[1]}$, $j^2_{\xp[1]}$, $\pi^1_{\xp[2]}$, $j^1_{\xp[2]}$, 
differ by a half integer for $\mathbf{x}\!=\!\mathbf{y}$, 
being zero otherwise. The explict expression of the 
matrix elements of the plaquette operator are:
\begin{eqnarray} 
&& \label{eq:PLAQUETTESU2}
  \langle \mathbf{\vec{\jmath}~'},\mathbf{\vec{\pi}'}| 
      U_{{\mathbf{y}},1,2} 
      | \mathbf{\vec{\jmath}},\mathbf{\vec{\pi}}\rangle =
\\  &&\nonumber  
 ~~~ = \frac{
    (-1)^{\sum_{i=1}^n \left( \left|\epsilon_i-\epsilon_{i\!+\!1}\right|
                     + \frac{ C^i_{\mathbf{y}} }{2} \right)}
  }{
    \sqrt{\prod_{i=1}^n 
          \left( 2 X^i_{\mathbf{y}}+ 1 \right) 
          \left( 2 Y^i_{\mathbf{y}}+ 1 \right) }
  }
\times
\\  &&\nonumber  
 ~~~ ~~~ \times  \prod_{i=1}^n ~ 
   R\left[{ \begin{array}{cc} 
            X^i_{\mathbf{y}} & X^{i\!+\!1}_{\mathbf{y}} \\  
            Y^i_{\mathbf{y}} & Y^{i\!+\!1}_{\mathbf{y}}   
    \end{array},C^i_{\mathbf{y}} }\right] 
\end{eqnarray}
where $\epsilon_i={X}^i_{\mathbf{y}}-{Y}^i_{\mathbf{y}}=\pm \frac{1}{2}$,  
$$  
\begin{array}{l}
{X}^1_{\mathbf{y}}\!=\!j_{\mathbf{x}}^1  ~, \\
{X}^2_{\mathbf{y}}\!=\!j_{\mathbf{x}}^2  ~, \\ 
{X}^3_{\mathbf{y}}\!=\!\pi^1_{\xp[2]}    ~, \\
{X}^4_{\mathbf{y}}\!=\!j^1_{\xp[2]}      ~, \\
{X}^5_{\mathbf{y}}\!=\!j^2_{\xp[1]}      ~, \\ 
{X}^6_{\mathbf{y}}\!=\!\pi^1_{\xp[1]}    ~,
\end{array}
~~
\begin{array}{l}
{Y}^1_{\mathbf{y}}\!=\!j_{\mathbf{x}}^{1\prime}  ~, \\
{Y}^2_{\mathbf{y}}\!=\!j_{\mathbf{x}}^{2\prime}  ~, \\ 
{Y}^3_{\mathbf{y}}\!=\!\pi^{1\prime}_{\xp[2]}    ~, \\
{Y}^4_{\mathbf{y}}\!=\!j^{1\prime}_{\xp[2]}      ~, \\
{Y}^5_{\mathbf{y}}\!=\!j^{2\prime}_{\xp[1]}      ~, \\ 
{Y}^6_{\mathbf{y}}\!=\!\pi^{1\prime}_{\xp[1]}    ~, 
\end{array}
~~
\begin{array}{l}
{C}^1_{\mathbf{y}}\!=\!\pi_{\mathbf{x}}^1     ~, \\
{C}^2_{\mathbf{y}}\!=\!j_{\xm[1\!+\!2]}^1     ~, \\
{C}^3_{\mathbf{y}}\!=\!j^2_{\xp[2]}           ~, \\
{C}^4_{\mathbf{y}}\!=\!\pi^1_{\xp[1\!+\!2]}   ~, \\
{C}^6_{\mathbf{y}}\!=\!j^1_{\xp[1]}           ~, \\
{C}^6_{\mathbf{y}}\!=\!j^2_{\xp[1\!-\!2]}      
\end{array}
$$
and $R\left[{ \begin{array}{cc} 
            X^i_{\mathbf{y}} & X^{i\!+\!1}_{\mathbf{y}} \\  
            Y^i_{\mathbf{y}} & Y^{i\!+\!1}_{\mathbf{y}}   
    \end{array},C^i_{\mathbf{y}} }\right] 
$ is equal to 
$$
  \textstyle  
  \sqrt{\! \frac{1-2 C^i_{\mathbf{y}}
               +X^i_{\mathbf{y}}
               +X^{i\!+\!1}_{\mathbf{y}} 
               +Y^i_{\mathbf{y}}
               +Y^{i\!+\!1}_{\mathbf{y}} 
     }{2} 
    \frac{3+2 C^i_{\mathbf{y}}
               +X^i_{\mathbf{y}}
               +X^{i\!+\!1}_{\mathbf{y}} 
               +Y^i_{\mathbf{y}}
               +Y^{i\!+\!1}_{\mathbf{y}} 
     }{2} \!\!
  }
$$
for $\left|\epsilon_i-\epsilon_{i\!+\!1}\right|=0$ and 
$$
\textstyle  
\sqrt{\! \frac{1+2 C^i_{\mathbf{y}}
             +X^i_{\mathbf{y}}
             -X^{i\!+\!1}_{\mathbf{y}} 
             +Y^i_{\mathbf{y}}
             -Y^{i\!+\!1}_{\mathbf{y}} 
          }{2} 
          \frac{1+2 C^i_{\mathbf{y}}
                 -X^i_{\mathbf{y}}
                 +X^{i\!+\!1}_{\mathbf{y}} 
                 -Y^i_{\mathbf{y}}
                 +Y^{i\!+\!1}_{\mathbf{y}} 
          }{2} \!\! 
    }
$$
for $\left|\epsilon_i-\epsilon_{i\!+\!1}\right|=1$

\section{CONCLUSIONS}
\protect\label{sec:conc}

In this work we have shown how {\it harmonic analysis on compact
Lie-Groups} together with {\it intertwining operators formalism} 
provide a useful basis of the Hilbert space of LGT and an
explicit expression for matrix elements of the Hamiltonian operator. 
Moreover, the latter can be expressed in terms of Wigner's $nJ$-symbols.
Such elements are well known for $SU(2)$
(see for example \cite{Yutsis:1962}). Explicit values for the 
$nJ$-symbol for unitary groups are found in \cite{ClebshesSUN}.

Our results are derived in terms of the knowledge of :
1) the full set of irreducible representations of the group,
2) a basis on the space of intertwining operators.
The generalization to gauge theories coupled to matter is
therefore straightforward.

The construction does not depend on the gauge group but only on its 
tensor category properties.

\section{ACKNOWLEDGMENTS} 

We thank E. Onofri and F. Di Renzo for helpful and 
enlightening discussions.  The work of R.D.P. is supported by a 
Dalla Riccia Fellowship. Partial support from grants CONACyT 
No. 3141P-E9608 and DGAPA-UNAM IN100397 is also acknowledged.


\end{document}